\documentclass[10pt]{article}
\usepackage{amssymb}
\usepackage{amscd}
\usepackage{latexsym}
\usepackage{boldtensors}
\usepackage{amsmath}
\usepackage{graphicx}
\newcommand{\newsec}{\setcounter{equation}{0}\section}%

\newcommand{\Rr}{{\mathbb R}}

\def\be{\begin{equation}}
\def\ee{\end{equation}}
\def\bea{\begin{eqnarray}}
\def\eea{\end{eqnarray}}

\def\d{{\,\rm d}}

\def\0{{\bf 0}}

\def\p{{\bf p}}

\def\x{{\bf x}}
\def\y{{\bf y}}

\def\P{{\bf P}}

\def\h2m{\frac{\hbar^2}{2m}}
\def\p0{{P_{\beta H^0_N}}}

\def\boldom{{~\omega}}
\newtheorem{theorem}{Theorem}[section]

\newtheorem{proposition}{Proposition}[section]

\begin{document}
\title{
\large\bf Self-interacting Brownian motion\footnote{Work supported by OTKA Grant No. K109577.}}
\author{Andr\'as S\"ut\H o\\Wigner Research Centre for Physics\\Hungarian Academy of Sciences\\P. O. B. 49, H-1525 Budapest, Hungary\\
E-mail: suto@szfki.hu\\}
\date{August 2, 2018}
\maketitle
\thispagestyle{empty}
\begin{abstract}
\noindent
We prove a property of Brownian bridges whose certain time-equidistant sequences of points are pairwise coupled by an interaction. Roughly saying, if the total time span $t$ of the bridge tends to infinity while the distance of its end points is fixed or increases slower than $\sqrt{t}$, the process asymptotically forgets this distance, just as in the absence of interaction. The conclusion remains valid if the bridge interacts in a similar way also with another set of trajectories. The main example for the interaction is the Coulomb potential.
\end{abstract}
\newsec{Introduction}
The problem we consider in this paper arises in Quantum Statistical Physics when a system of interacting particles is investigated by utilizing the Feynman-Kac formula [F1-2, G1-4]. It is in relation with Bose-Einstein condensation that we do not discuss here. Let $P^\beta_{\0\x}(\d\boldom)$ denote the Wiener measure for Brownian paths $\boldom:[0,\beta]\rightarrow\Rr^\nu$ with $\boldom(0)=\0$, $\boldom(\beta)=\x$. It is generated by the Gaussian functions
\be
\psi_t(\x)=\lambda_t^{-\nu}e^{-\pi \x^2/\lambda_t^2},\quad 0\leq t\leq\beta.
\ee
In physics $\lambda_t$ is the thermal de Broglie wave length at inverse temperature $t$, $\lambda_t=\sqrt{2\pi\hbar^2 t/m}$, where $m$ is the mass of the particle. In particular,
\be\label{norms}
\int P^\beta_{\0\x}(\d\boldom)=\lambda_\beta^{-\nu}e^{-\pi\x^2/\lambda_\beta^2}.
\ee
Our starting observation is that
\be
\lim_{\x^2/\lambda_\beta^2\to\,0}\frac{\int P^\beta_{\0\x}(\d\boldom)}{\int P^\beta_{\0\0}(\d\boldom)}=1,
\ee
and we wish to prove a similar result in the case of interaction. For this we take a real function $u(\x)$ on $\Rr^\nu$ that depends only on $|\x|$; the interaction between points $\x$ and $\y$ is $u(\x-\y)$. The time span of the bridge is $n\beta$, where $\beta>0$ and $n$ is a positive integer. Given a trajectory $\boldom:[0,n\beta]\rightarrow\Rr^\nu$, its self-interaction is defined as
\be
U(\boldom)=\frac{1}{\beta}\sum_{0\leq k<l\leq n-1}\int_0^\beta u(\boldom(l\beta+t)-\boldom(k\beta+t))\d t,
\ee
and our interest is to give a lower bound on
\be
\int P^\beta_{\0\x}(\d\boldom)e^{-\beta U(\boldom)}\left/\int P^\beta_{\0\0}(\d\boldom)e^{-\beta U(\boldom)}\right..
\ee
As to the physical background, let
\be
H_n=-\frac{\hbar^2}{2m}\sum_{i=1}^n\Delta_{\x_i}+\sum_{1\leq i<j\leq n} u(\x_j-\x_i),
\ee
the energy operator of $n$ interacting particles of mass $m$. For suitable choices of $u$, $e^{-\beta H_n}$ has an integral kernel,
and by the Feynman-Kac formula we have
\be
\int
e^{-\beta H_n}(\0,\x_2,\ldots,\x_n;\x_2,\ldots,\x_n,\x)
\d\x_2\ldots\d\x_n =\int P^{n\beta}_{\0\x}(\d\boldom)e^{-\beta U(\boldom)}.
\ee

In addition, we shall consider also the problem when, apart from a Brownian bridge $\boldom_0$ of time span $n\beta$ and self-interaction $U_1$ generated by a pair potential $u_1$, there is also a set of $M$ other trajectories $\boldom_i:[0,\beta]\rightarrow\Rr^\nu$
\be
\boldom^{M}=(\boldom_1,\ldots,\boldom_{M}),\qquad \boldom_i(0)=\x_i,\quad \boldom(\beta)=\y_i.
\ee
These may interact among themselves; their potential energy is then
\be
U_2(\boldom^M)=\frac{1}{\beta}\sum_{1\leq i<j\leq M}\int_0^\beta u_2(\boldom_i(t)-\boldom_j(t))\d t,
\ee
with a pair potential $u_2$. The inclusion of this interaction is optional; what is important is an interaction between $\boldom_0$ and $\boldom^M$,
\be\label{external}
V\left(\boldom_0,\boldom^{M}\right)=\frac{1}{\beta}\sum_{k=0}^{n-1}\sum_{i=1}^{M}\int_0^\beta u_3(\boldom_0(k\beta+t)-\boldom_i(t))\d t,
\ee
acting as an external field for $\boldom_0$. Both $u_2(\x)$ and $u_3(\x)$ depend only on $|\x|$. The underlying energy operator in this case is that of $n$ particles of mass $m_1$ with pair interaction $u_1$ and $M$ particles of mass $m_2$ with pair interaction $u_2$, the two sets pairwise interacting via $u_3$:
\be\label{Ham-n,M}
H_{n,M}=-\frac{\hbar^2}{2m_1}\sum_{i=1}^n\Delta_{\x_i}+\sum_{1\leq i<j\leq n} u_1(\x_j-\x_i) +\sum_{i=1}^n\sum_{j=1}^M u_3(\x_i-\y_j) -\frac{\hbar^2}{2m_2}\sum_{i=1}^M\Delta_{\y_i}+\sum_{1\leq i<j\leq M} u_2(\y_j-\y_i).
\ee
Some special cases of the external field will be presented in Section 3.

\newsec{Isolated Brownian motion}
Here we treat a Brownian bridge of time span $n\beta$ under self-interaction alone.

\begin{theorem}\label{isolated}
Let $u$ be twice differentiable outside $\0$. If $\Delta u\leq 0$, then
\be\label{ratio}
\lim_{n\to\infty,\,\x^2/n\to\,0}\frac{\int P^{n\beta}_{\0\x}(\d\boldom)e^{-\beta U(\boldom)}}{\int P^{n\beta}_{\0\0}(\d\boldom)e^{-\beta U(\boldom)}}\geq 1.
\ee
\end{theorem}


\vspace{10pt}
\noindent
{\em Proof.} We start with the identity
\be
\int P^{n\beta}_{\0\x}(\d\boldom)\phi(\boldom)=\exp\left\{-\frac{\pi \x^2}{n\lambda_\beta^2}\right\} \int P^{n\beta}_{\0\0}(\d\boldom)\phi(\widetilde{\boldom})
\ee
where
\be
\widetilde{\boldom}(t)=\boldom(t)+\frac{t}{n\beta}\x,
\ee
see [G1]. Applying this to $\phi(\boldom)=\exp-\beta U(\boldom)$ and using the notation
\be
E_\boldom(\x)=\sum_{0\leq k<l\leq n-1}\int_0^\beta u\left(\boldom(l\beta+t)-\boldom(k\beta+t)+\frac{l-k}{n}\x\right)\d t
\ee
we obtain
\be
\int P^{n\beta}_{\0\x}(\d\boldom)e^{-\beta U(\boldom)}
=\exp\left\{-\frac{\pi \x^2}{n\lambda_\beta^2}\right\} \int P^{n\beta}_{\0\0}(\d\boldom)
e^{-E_\boldom(\x)}.
\ee
Now
\be
\lim_{n\to\infty,\,\x^2/n\to\,0}\exp\left\{-\frac{\pi \x^2}{n\lambda_\beta^2}\right\}=1,
\ee
so for
\be
I^{n\beta}(\x):=\int P^{n\beta}_{\0\0}(\d\boldom)e^{-E_\boldom(\x)}
\ee
we must prove that
\be
\lim_{n\to\infty,\,\x^2/n\to\,0}\frac{I^{n\beta}(\x)}{I^{n\beta}(\0)}\geq 1.
\ee
Actually, the stronger result
\be
\frac{I^{n\beta}(\x)}{I^{n\beta}(\0)}\geq 1,
\ee
without taking the limit, also holds. Note that $I^{n\beta}(\x)$ is a spherical function, thus it has an extremum at $\0$. We show that $I^{n\beta}(\0)$ is a global minimum by proving that $\Delta I^{n\beta}(\x)\geq 0$ everywhere. Indeed, this implies that $I^{n\beta}$ is a convex even function of $\x$ along every axis passing through $\0$. Now
\be
~\nabla I^{n\beta}(\x)=-\int P^{n\beta}_{\0\0}(\d\boldom)e^{-E_\boldom(\x)}~\nabla E_\boldom(\x),
\ee
\be
\Delta I^{n\beta}(\x)=\int P^{n\beta}_{\0\0}(\d\boldom)e^{-E_\boldom(\x)} \left[\left|~\nabla E_\boldom(\x)\right|^2-\Delta E_\boldom(\x)\right].
\ee
So $\Delta I^{n\beta}(\x)\geq 0$ if $\Delta E_\boldom(\x)\leq 0$, and the result follows from
\bea\label{Delta-u}
\Delta E_\boldom(\x) &=& \sum_{k<l}\int_0^\beta\Delta_\x u\left(\boldom(l\beta+t)-\boldom(k\beta+t)+\frac{l-k}{n}\x\right)\d t \nonumber\\ &=& \sum_{k<l}\left(\frac{l-k}{n}\right)^2\int_0^\beta\Delta u\left(\boldom(l\beta+t)-\boldom(k\beta+t)+\frac{l-k}{n}\x\right)\d t.
\eea

The general solution of $\Delta u\leq 0$ is easily obtained. Define $f:[0,\infty)\rightarrow \Rr\cup\{\infty\}$ by $u(\y)=f(\y^2)$. Because
\be
\Delta u(\y)=2\left[\nu f'(\y^2)+2\y^2 f''(\y^2)\right],
\ee
one must find those $f$ satisfying
\be\label{cond1}
\nu f'(s) +2s f''(s)\leq 0\quad\mbox{for any $s> 0$}.
\ee
Let $g(s)$ be a nonnegative function, and consider the differential equation
\be\label{int-eq}
\nu f'(s) +2s f''(s)=-g(s).
\ee
Choose some $a,b>0$, then for the boundary conditions
\be
f(a)=c_1,\qquad f'(b)=c_2 b^{-\nu/2}
\ee
the solution of Eq.~(\ref{int-eq}) is
\be
f(s)=c_1-\frac{1}{2}\int_a^s \d x\, x^{-\nu/2} \int_b^x \d t\, g(t)t^{\nu/2-1} +c_2\left\{\begin{array}{cc}
\frac{s^{1-\nu/2}-a^{1-\nu/2}}{1-\nu/2}, & \nu\neq 2\\
(\ln s-\ln a), & \nu=2.
\end{array}\right.\qquad\Box
\ee

\vspace{10pt}
\noindent
{\bf Examples.}
For $g=0$ we can obtain the $\nu$-dimensional Coulomb potential of both signs. In general, $f(s)=s^{-\alpha}$ is a solution of Eq.~(\ref{int-eq}) for $g(s)=\alpha(\nu-2-2\alpha)s^{-\alpha-1}$ which is nonnegative if $\alpha\leq\nu/2-1$. Another interesting solution is obtained in dimensions $\nu\geq 3$ if
\be
a=b=+\infty,\quad c_1=c_2=0,\quad g(s)=s^{-\alpha},\quad\alpha>\nu/2>1.
\ee
It reads
\be
f(s)=\frac{-s^{-\alpha+1}}{(\alpha-1)(2\alpha-\nu)}\quad\mbox{or}\quad u(\x)=\frac{-|\x|^{-2\alpha+2}}{(\alpha-1)(2\alpha-\nu)}.
\ee
In three dimensions with $\alpha=4$ this is the form of the induced dipole-dipole interaction which appears in the Lennard-Jones potential.

\newsec{Self-interacting Brownian bridge in an external field}

The external field is created by a set of other trajectories, and is given by Eq.~(\ref{external}).
Let
\bea
E_{\boldom_0}(\x)=\sum_{0\leq k<l\leq n-1}\int_0^\beta u_1\left(\boldom_0(l\beta+t)-\boldom_0(k\beta+t)+\frac{l-k}{n}\x\right)\d t,\nonumber\\
E_{\boldom_0,\boldom^{M}}(\x)=\sum_{k=0}^{n-1}\sum_{i=1}^{M}\int_0^\beta u_3\left(\boldom_0(k\beta+t)-\boldom_i(t)+\frac{k\beta+t}{n\beta}\x\right)\d t,
\eea
and
\be
I^{n\beta}_{\boldom^{M}}(\x)=\int P^{n\beta}_{\0\0}(\d\boldom_0)e^{-E_{\boldom_0}(\x)-E_{\boldom_0,\boldom^{M}}(\x)-\beta U_2(\boldom^M)}.
\ee
As before,
\be\label{separate-Gauss}
\int P^{n\beta}_{\0\x}(\d\boldom_0)e^{-\beta\left[U_1(\boldom_0)+ V\left(\boldom_0,\boldom^{M}\right)+U_2\left(\boldom^M\right)\right]} =e^{-\pi\x^2/\lambda_{n\beta}^2}I^{n\beta}_{\boldom^{M}}(\x),
\ee
where $\lambda_{n\beta}=\sqrt{2\pi\hbar^2 n\beta/m_1}$.

\begin{proposition}\label{Delta-I}
If $u_1$ and $u_3$ are twice differentiable outside the origin and $\Delta u_1\leq 0$, $\Delta u_3\leq 0$, then $\Delta I^{n\beta}_{\boldom^{M}}\geq 0$.
\end{proposition}

\vspace{10pt}
\noindent
{\em Proof.} Again, a direct computation gives
\bea
\Delta I^{n\beta}_{\boldom^{M}}(\x) &=& \int P^{n\beta}_{\0\0}(\d\boldom_0)e^{-E_{\boldom_0}(\x)-E_{\boldom_0,\boldom^{M}}(\x)-\beta U_2\left(\boldom^M\right)}\nonumber\\
&\times& \left[\left|~\nabla E_{\boldom_0}(\x)+~\nabla E_{\boldom_0,\boldom^M}(\x)\right|^2-\Delta E_{\boldom_0}(\x)-\Delta E_{\boldom_0,\boldom^M}(\x)\right].
%
\eea
Recalling the identity (\ref{Delta-u}) and
\be
\Delta E_{\boldom_0,\boldom^M}(\x) =\sum_{k=0}^{n-1}\sum_{i=1}^{M}\int_0^\beta\left(\frac{k\beta+t}{n\beta}\right)^2 \Delta u_3\left(\boldom_0(k\beta+t)-\boldom_i(t)+\frac{k\beta+t}{n\beta}\x\right)\d t
\ee
the result follows.$\quad\Box$

\vspace{15pt}
$I^{n\beta}_{\boldom^{M}}(\x)$ is not a spherical function, so it may not be minimal at $\x=\0$. If, for example,
\be
\mathrm{dist}\left(\x,\{\x_i\}_{i=1}^M\right) \ll\mathrm{dist}\left(\0,\{\x_i\}_{i=1}^M\right)
\ee
then possibly
\be
E_{\boldom_0,\boldom^{M}}(\x)\gg E_{\boldom_0,\boldom^{M}}(\0)
\ee
and therefore $I^{n\beta}_{\boldom^{M}}(\x)<I^{n\beta}_{\boldom^{M}}(\0)$.

Let ${\cal S}\left(I^{n\beta}_{\boldom^{M}}\right)(\x)$ denote a spherical symmetrization of $I^{n\beta}_{\boldom^{M}}(\x)$, obtained by integration over a set of $\boldom^M$. Examples will be given below. As a function of $\x$, ${\cal S}\left(I^{n\beta}_{\boldom^{M}}\right)(\x)$ depends only on $|\x|$, and may preserve some dependence on $\boldom^M$. From Eq.~(\ref{separate-Gauss}),
\be
{\cal S}\left(\int P^{n\beta}_{\0\x}(\d\boldom_0)e^{-\beta\left[U_1(\boldom_0)+ V\left(\boldom_0,\boldom^{M}\right)+U_2\left(\boldom^M\right)\right]}\right)(\x) =e^{-\pi\x^2/\lambda_{n\beta}^2}{\cal S}\left(I^{n\beta}_{\boldom^{M}}\right)(\x).
\ee
One can differentiate inside the integral over $\boldom^M$, so
\be
\Delta{\cal S}\left(I^{n\beta}_{\boldom^{M}}\right)(\x)={\cal S}\left(\Delta I^{n\beta}_{\boldom^{M}}\right)(\x).
\ee

\begin{theorem}
For $u_1$ and $u_3$ as in Proposition \ref{Delta-I},
\be
\lim_{n\to\infty,\x^2/n\to 0}\frac{{\cal S}\left(\int P^{n\beta}_{\0\x}(\d\boldom_0)e^{-\beta\left[U_1(\boldom_0)+ V\left(\boldom_0,\boldom^{M}\right)+U_2\left(\boldom^M\right)\right]}\right)(\x)}{{\cal S}\left(\int P^{n\beta}_{\0\x}(\d\boldom_0)e^{-\beta\left[U_1(\boldom_0)+ V\left(\boldom_0,\boldom^{M}\right)+U_2\left(\boldom^M\right)\right]}\right)(\0)}\geq 1.
\ee
\end{theorem}

\vspace{10pt}
\noindent
{\em Proof.}
$\Delta I^{n\beta}_{\boldom^{M}}\geq 0$ implies $\Delta{\cal S}\left(I^{n\beta}_{\boldom^{M}}\right)\geq 0$ and, because ${\cal S}\left(I^{n\beta}_{\boldom^{M}}\right)(\x)$ is spherical,
$${\cal S}\left(I^{n\beta}_{\boldom^{M}}\right)(\x)\geq {\cal S}\left(I^{n\beta}_{\boldom^{M}}\right)(\0).$$
Then,
\be
\lim_{n\to\infty,\x^2/n\to 0}\frac{{\cal S}\left(\int P^{n\beta}_{\0\x}(\d\boldom_0)e^{-\beta\left[U_1(\boldom_0)+ V\left(\boldom_0,\boldom^{M}\right)+U_2\left(\boldom^M\right)\right]}\right)(\x)}{{\cal S}\left(\int P^{n\beta}_{\0\x}(\d\boldom_0)e^{-\beta\left[U_1(\boldom_0)+ V\left(\boldom_0,\boldom^{M}\right)+U_2\left(\boldom^M\right)\right]}\right)(\0)} =\lim_{n\to\infty,\x^2/n\to 0}\frac{{\cal S}\left(I^{n\beta}_{\boldom^{M}}\right)(\x)}{{\cal S}\left(I^{n\beta}_{\boldom^{M}}\right)(\0)}    \geq 1.
\ee
$\quad\Box$

\vspace{10pt}
\noindent
{\bf Examples.}\\
(i) Fix some $L>0$ and define
\be
{\cal S}\left(I^{n\beta}_{\cdot}\right)(\x)=\sum_{\pi\in S_{M}}\prod_{i=1}^{M}\int_{|\x_i|<L}\d\x_i\int P^\beta_{\x_i\x_{\pi(i)}}(\d\boldom_i)I^{n\beta}_{\boldom^{M}}(\x)
\ee
where $S_{M}$ is the set of permutations of $\{1,\ldots,M\}$. This is the typical choice when the particles associated with $\boldom^M$ are bosons. The physically interesting case is $n\propto M\propto L^\nu$ and $n\to\infty$ replaced by $L\to\infty$. If $n\propto L^\nu$, for $\x^2/n\to 0$ to hold one must have $|\x|=o(L^{\nu/2})$. In $\nu\geq 3$ dimensions all $\x$ of length $|\x|=O(L)$ satisfy this condition. When $n=M$, $u_1=u_2=-u_3=u$ with $\Delta u=0$, the system is a neutral two-component plasma.\\
(ii) The second set of particles can be treated classically. This is the $m_2=\infty$ limit in which both the kinetic and the potential energy of these particles (the last two terms of the Hamiltonian (\ref{Ham-n,M})) can be dropped. Now $\boldom_i(t)\equiv\x_i$, $\boldom^M=\x^M=(\x_1,\ldots,\x_M)$, and the interaction between the two sets becomes
\be
V\left(\boldom_0,\x^{M}\right)=\frac{1}{\beta}\sum_{k=0}^{n-1}\sum_{i=1}^{M}\int_0^\beta u_3(\boldom_0(k\beta+t)-\x_i)\d t.
\ee
Spherical symmetrization can be done by keeping all $|\x_i-\x_j|$ unchanged. Let $g$ denote a general element of the rotation group $SO(\nu)$, and let $\mu$ be the Haar measure on $SO(\nu)$. Define
\be
{\cal S}\left(I^{n\beta}_{(\x_1,\ldots,\x_M)}\right)(\x)=\int I^{n\beta}_{(g\x_1,\ldots,g\x_M)}(\x)\,\mu(\d g).
\ee
Because $u_1$ and $u_3$ are spherical functions and the measure $P^{n\beta}_{\0\0}$ is rotation invariant, one actually has
\be
{\cal S}\left(I^{n\beta}_{(\x_1,\ldots,\x_M)}\right)(\x)=\int I^{n\beta}_{(\x_1,\ldots,\x_M)}(g\x)\,\mu(\d g).
\ee
When $n=M$, $u_1=-u_3=u$ with $\Delta u=0$, we have a neutral system of light charged particles moving in the field of immobile heavy ions.


\vspace{20pt}
\noindent{\Large\bf References}
\begin{enumerate}
\item[{[Fe1]}] Feynman R. P.: {\em Space-time approach to non-relativistic quantum mechanics.} Rev. Mod. Phys. {\bf 20}, 367-387 (1948).
\item[{[Fe2]}] Feynman R. P.: {\em Atomic theory of the $\lambda$ transition in helium.} Phys. Rev. {\bf 91}, 1291-1301 (1953).
\item[{[G1]}] Ginibre J.: {\em Some applications of functional integration in Statistical Mechanics.} In: {\em Statistical Mechanics and Quantum Field Theory}, eds. C. De Witt and R. Stora, Gordon and Breach (New York 1971).
\item[{[G2]}] Ginibre J.: {\em Reduced density matrices of quantum gases. I. Limit of infinite volume.} J. Math. Phys. {\bf 6}, 238-251 (1965).
\item[{[G3]}] Ginibre J.: {\em Reduced density matrices of quantum gases. II. Cluster property.} J. Math. Phys. {\bf 6}, 252-262 (1965).
\item[{[G4]}] Ginibre J.: {\em Reduced density matrices of quantum gases. III. Hard-core potentials.} J. Math. Phys. {\bf 6}, 1432-1446 (1965).
\end{enumerate}

\end{document}